\documentclass[conference]{IEEEtran}
\IEEEoverridecommandlockouts
\usepackage{cite}
\usepackage{amsmath,amssymb,amsfonts}
\usepackage{algorithmic}
\usepackage{graphicx}
\usepackage{textcomp}
\usepackage{xcolor}
\usepackage{url}
\usepackage{xspace}
\usepackage{balance}

\def\BibTeX{{\rm B\kern-.05em{\sc i\kern-.025em b}\kern-.08em
    T\kern-.1667em\lower.7ex\hbox{E}\kern-.125emX}}
\def\sysname{\texttt{DAT}\xspace}
\begin{document}

\title{Object Model Analysis of a Supercomputer with Digital Twin}

\author{
\IEEEauthorblockN{
Shilpika$^{1,*}$,
George K.~Thiruvathukal$^{1,2,*}$,
Venkatram Vishwanath$^{1}$,
and Michael E.~Papka$^{1,3}$
}
\IEEEauthorblockA{
$^{1}$Leadership Computing Facility,
Argonne National Laboratory,
Argonne, IL, USA
}
\IEEEauthorblockA{
$^{2}$Department of Computer Science,
Loyola University Chicago,
Chicago, IL, USA
}
\IEEEauthorblockA{
$^{3}$Department of Computer Science,
University of Illinois at Chicago,
Chicago, IL, USA
}
\IEEEauthorblockA{
$^{*}$ These authors contributed equally to this work.
}
}

\maketitle

\begin{abstract}
Operators and developers need a mental model of both the structure and the live behavior of a large supercomputer, but its physical layout, logical organization, and streams of per-node telemetry are difficult to relate to one another, making it hard to trace a metric or event back to a specific hardware component. We present \sysname, an interactive three-dimensional digital analytics twin of a compute cluster built in a real-time game engine, Unreal Engine. \sysname expands a compact, parametric description of a supercomputer---a reusable Digital Twin Prototype (DTP)---into a navigable Digital Twin Instance (DTI) that mirrors its physical containment hierarchy of racks, chassis, blades, and network links, encoding each node's role and health in its appearance, while a lightweight event-driven simulator animates job and hardware activity over a virtual clock. Our current implementation adds a two-path node-selection mechanism, unifying direct 3D pointing with command-shell queries, that opens an in-world visual-analytics panel beside any selected component showing summary statistics and live, time-varying metrics. We describe this architecture, report qualitative behavior from the working prototype, and outline the path toward driving the panels with recorded telemetry and in-situ anomaly detection.
\end{abstract}


\balance

\section{Introduction}
Operators, facility staff, and application developers need a shared mental
model of a supercomputer's behavior, but the information they must integrate
spans three different representations: a spatial \emph{physical} view
(cabinets, chassis, blades, and nodes), a \emph{logical} view of roles
(compute, storage, gateway, service, user-access, switch), and a
\emph{dynamic} view of per-node telemetry, job events, and hardware events
that change in real time. Traditional dashboards present the dynamic view
well but detach it from the spatial one, while floor diagrams capture the
spatial view but stay static, so relating a metric spike back to a specific
component and its neighbors remains a challenging, error-prone task.

A digital twin---a live, navigable model mirroring both structure and
behavior---is an appealing way to unify these views, and game engines are
attractive vehicles for one: real-time rendering, spatial navigation, and
asset pipelines come out of the box. The open question is how to display
surface-dense per-node information \emph{in place}, without overwhelming the
scene or falling back to a separate two-dimensional dashboard.

\begin{figure}[h]
\centerline{\includegraphics[width=\columnwidth]{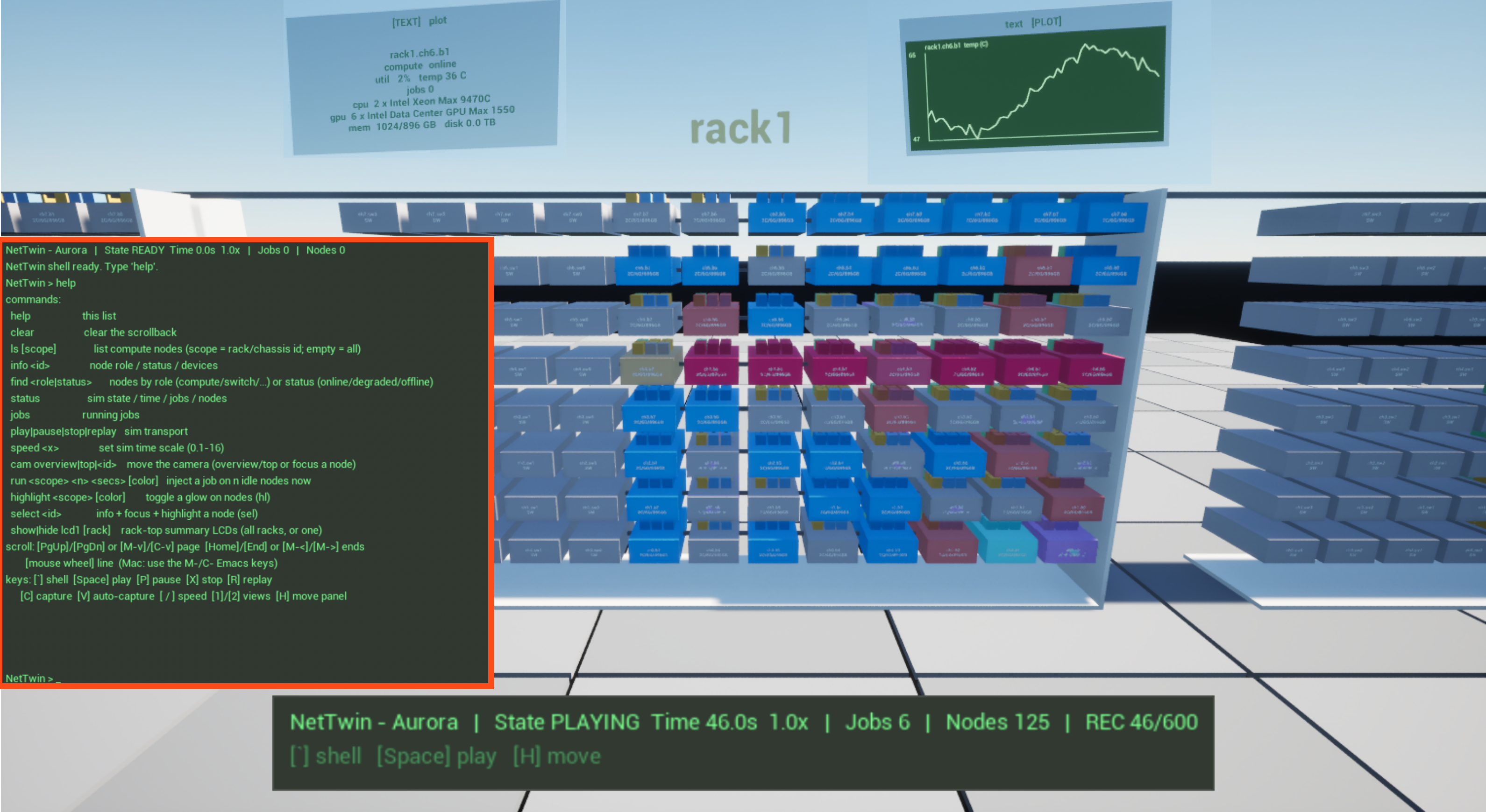}}
\caption{An example of \sysname simulation with three racks captured automatically from within the game.
The nodes as cubes are colored based on which job is running on them.
The in-world LCD panel attached to a selected node.
The panel billboards toward the camera and shows the node's identifier, role, and status, live metrics, and device specification.
The interactive command shell with the available options is shown in a red outline.}
\label{fig:lcd}
\end{figure}

This paper reports progress on \sysname, a digital twin of a compute
cluster in Unreal Engine that already expands a parametric system description
into a navigable scene (see Fig.~\ref{fig:lcd}), encodes node role and health visually, and animates
job and hardware activity over a virtual clock.


Our current work adds the
missing link between the spatial and dynamic views: an interactive, in-world
display attached to any node or rack. Concretely, our contributions are:

\begin{itemize}

 \item a metadata-driven Digital Twin Prototype (DTP) that spawns a Digital
       Twin Instance (DTI) per description, for a lightweight, reusable
       architecture;
  \item a two-path node-selection mechanism unifying direct 3D pointing with
        symbolic selection from a command shell (CLI), routed through a single
        selection state;
  \item an in-world 2D display beside the selected node or rack that
        billboards toward the camera and refreshes live at a fixed cadence;
        and
  \item a per-node and per-rack metric layer that synthesizes plausible,
        time-varying values today so recorded telemetry can be substituted
        transparently later, with video capture for tracking anomalous
        events.
\end{itemize}
We describe the supporting architecture, report implementation status and
qualitative observations, and discuss trade-offs and the roadmap toward real
telemetry and in-situ plotting.

\section{Related Work}

The growth of exascale HPC and large AI data centers has driven multi-physics modeling frameworks that optimize energy efficiency, predict power swings, and manage system operations.
Related work spans several categories: digital twin maturity models, thermo-fluidic cooling simulation, network congestion modeling, job scheduling optimization, and immersive visual analytics.
These paradigms establish the context against which we position our proposed \sysname framework.

\subsection{High-Level Summary of ExaDigiT}

ExaDigiT is an open-source effort to build cyber-physical digital twins of liquid-cooled supercomputers and utility plants, modeled after Frontier at Oak Ridge \cite{brewer2024exadigit, holmen2024exascale}.
Its architecture couples three modules: the Python-based Resource Allocator and Power Simulator (RAPS), which simulates scheduling and power draw including AC-to-DC/SIVOC losses \cite{brewer2024exadigit, wojda2024dynamic}; a Modelica-based thermo-fluidic model solving mass, pressure, and energy conservation from cooling towers to rack-level units \cite{kumar2024thermo, greenwood2024thermo}; and a visual analytics layer on Unreal Engine 5 for real-time monitoring and what-if exploration \cite{maiterth2024visualizing, b2}.
Workloads are ingested as CPU/GPU traces and translated into thermal loads driving the solvers \cite{brewer2024exadigit}.

\subsection{Hierarchical Digital Twin Maturity and Standards}

Datacenter digital twin design follows a five-level maturity hierarchy: descriptive twins statically replicate physical assets; informative twins add real-time metric streams; predictive twins add data-driven forecasting; comprehensive twins use first-principles solvers for what-if engineering; and autonomous twins reach SOTA via reinforcement learning and control loops \cite{brewer2024exadigit, b1}.
Building these environments requires interoperability standards such as SSP and the Functional Mock-up Interface (FMI), exchanging models as Functional Mock-up Units (FMUs) \cite{rosenlund2024objectively}.
To ensure credibility, researchers represent legacy domains as convex hulls---mapping regions such as altitude/Mach or coolant enthalpy/mass flow---to verify compatibility before co-simulation \cite{rosenlund2024objectively}.

\subsection{Thermal and Cooling System Modeling}

Modeling thermodynamics and fluid dynamics in liquid-cooled supercomputers means balancing accuracy with real-time execution. Traditional CFD packages, such as 6SigmaDC or 6SigmaRoom, solve Navier-Stokes equations for micro-level airflow and temperature, but their complexity precludes system-level simulation \cite{brewer2024exadigit}. System-level simulators instead use acausal solvers---the open-source TRANSFORM and Modelica Buildings libraries compiled in Dymola or OpenModelica---to model multi-loop facility dynamics in real-time \cite{kumar2024thermo, greenwood2024thermo}. Because these solvers still carry overhead, benchmarks such as LC-Opt and DCRL-Green explore deep learning surrogates---PINNs, DeepONet, or Fourier Neural Operators---to accelerate thermal-hydraulic updates for real-time reinforcement learning control \cite{naug2025lcopt, nigatu2026foundations}.

\subsection{Network Simulation and Congestion Analytics}

Network digital twin architectures model either the control plane or the data plane. VM-based emulation platforms, such as NVIDIA Air or NetGraph, virtualize network operating systems and switch configurations to validate routing and security policies before deployment \cite{brewer2024exadigit, holmen2024exascale}. Data-plane simulators instead study packet-forwarding: cycle-accurate tools such as the Structural Simulation Toolkit (SST Macro) ingest parallel MPI traces from crayfish profiling, Darshan, or SST DUMPI to simulate routing, latency, and inter-job interference across topologies like Frontier's 151,885-link Slingshot Dragonfly network \cite{holmen2024exascale}. Congestion events are aggregated into coarse-grained intervals, supported by collectors like the Lightweight Distributed Metric Service (LDMS) for system-wide correlation \cite{holmen2024exascale, b4}.

\subsection{Job Scheduling and Resource Management}

Batch scheduling determines resource utilization and queue turnaround in large-scale HPC systems. Standalone simulators such as FastSim, Slurm Simulator, Batsim, and CQSim evaluate scheduling algorithms (FCFS, EASY backfill, priority queuing) in isolation, assuming the infrastructure operates in a thermodynamic vacuum \cite{maiterth2025hpc}. To capture the physical consequences of scheduling, ExaDigiT's Scheduled-RAPS (S-RAPS) couples scheduling events with power and cooling simulations \cite{maiterth2025hpc}. This lets researchers replay open telemetry datasets to simulate how scheduling drives multi-megawatt power swings and thermal imbalances, informing energy-aware, carbon-aware, and ML-guided strategies \cite{maiterth2025hpc}.

\subsection{Immersive Visual Analytics and Desktop Dashboards}

A digital twin's visual representation is the primary interface for monitoring, anomaly detection, and planning. Descriptive twins typically use photorealistic engines like NVIDIA Omniverse or Unreal Engine 5, with Niagara particle systems to visualize network topologies \cite{brewer2024exadigit, maiterth2024visualizing, b2}. Immersive tools like TwinVar instead use AR/VR HMDs to overlay metric widgets onto physical or virtual hardware \cite{maiterth2024visualizing, b2}. While traditional architectures split 3D visualization from separate 2D graphs, modern exascale twins embed interactive dashboards (Grafana, Apache Druid) within the virtual environment, letting operators run what-if simulations, inspect cores, and query scheduler states without leaving the immersive view \cite{brewer2024exadigit, maiterth2024visualizing}.

Taken together, these efforts push toward either high-fidelity physical replication or a purely logical simulation of scheduling, cooling, and network behavior, each at the cost of complexity that can obscure day-to-day operational understanding \cite{maiterth2024visualizing, holmen2024exascale}. \sysname instead treats simulation as a means rather than an end: it keeps just enough physical grounding---containment, topology, and role---to make every metric traceable to a place on the machine-room floor, while representing behavior at a coarser, discrete-event level rather than a first-principles one \cite{brewer2024exadigit, maiterth2024visualizing}. This physically-anchored, intentionally lightweight design favors everyday comprehension over engineering-grade prediction, motivating the architecture we describe next.

\section{System Overview}
\sysname is written in C++ against a real-time game engine. The design separates
a pure data model from the actors that render it and from the simulator that
drives activity, so that each concern can evolve independently. This section
summarizes the existing platform and then details the interactive display.

\begin{figure}[t]
\centerline{\includegraphics[width=\columnwidth]{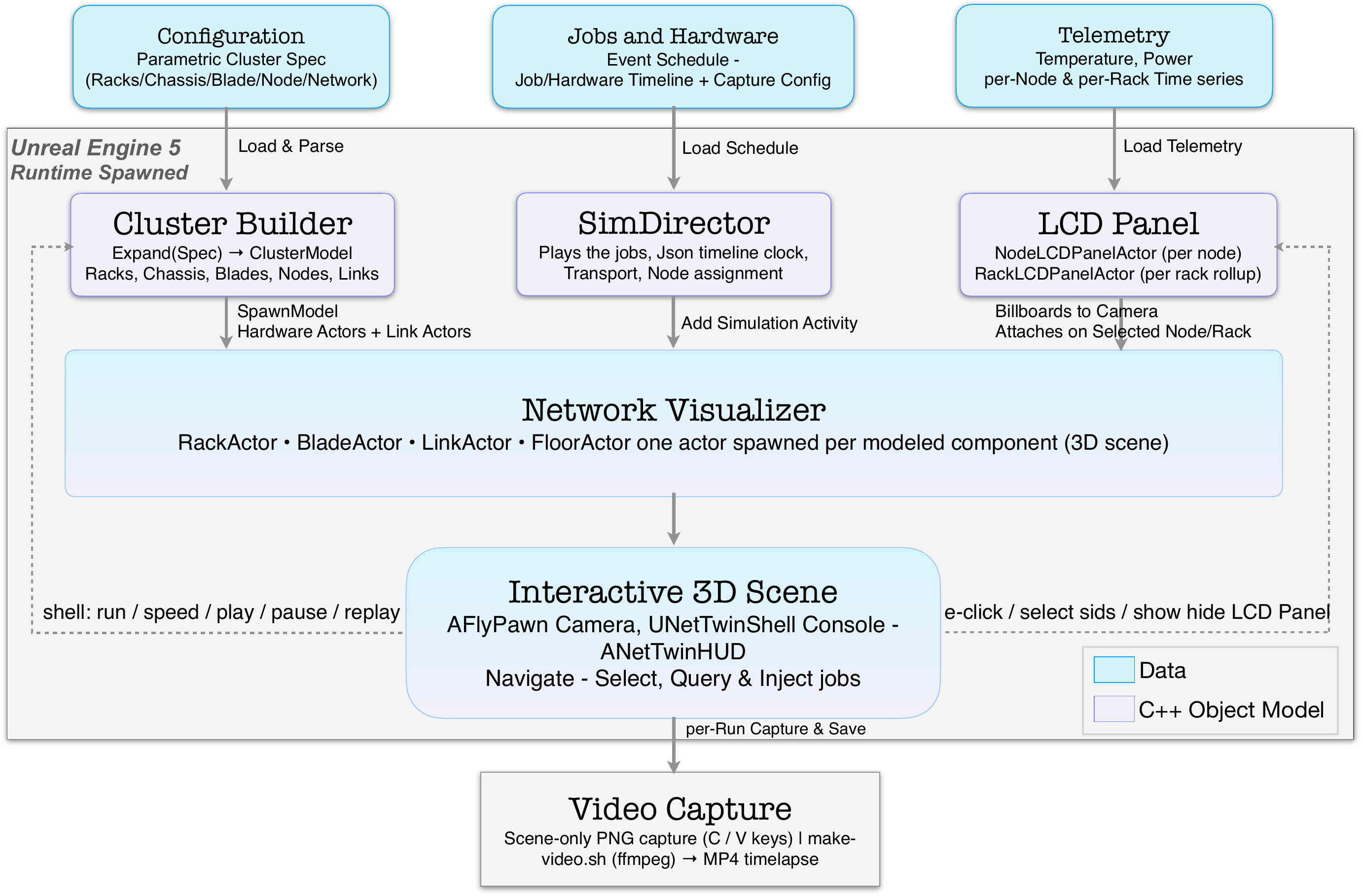}}
\caption{System Overview of our Parametric Digital Analytics Twin Architecture}
\label{fig:sysoverview}
\end{figure}

\subsection{Metadata-Driven Data Model} 

Fig.~\ref{fig:sysoverview} shows the architecture diagram of our parametric \sysname.
The model is a compact, reflected description of a system, deliberately
capturing organizing principles drawn from a modern exascale-class architecture.
It represents recursive physical containment: a system contains racks, a rack
contains chassis, and a chassis contains blades, through a parent identifier on
each node. A node's \emph{role} (compute, storage, gateway, service,
user-access, or switch) is kept orthogonal to its \emph{health status} (online,
degraded, or offline), because the two vary for different reasons. Each node
also carries a device-capability \emph{summary}: processor and accelerator
counts and models, system and high-bandwidth memory capacities, local storage,
and fabric interface count. Typed links between nodes record interconnect kind,
per-link bandwidth, and bundle multiplicity. The model is intentionally a
summary for visualization rather than a simulation of internal microarchitecture, therefore resulting in a DTP from which one can instantiate DTI’s. Each DTI is a single physical containment including a supercomputer.

Rather than author every node by hand, a builder expands a small specification
into the full model. This keeps large systems tractable to describe and lets us
render a configurable subset (for example, a handful of cabinets) while
preserving identifiers, so that the same names work whether one cabinet or the
whole floor is materialized.

\subsection{Scene Construction and Visual Encoding}
A visualizer (ClusterBuilder, SimDirector, LCD Panel, to Network Visualizer) consumes the expanded model and lays the scene out on a virtual
machine-room floor: racks as cabinets, chassis as rows, and blades as a grid on
the cabinet face, with local and global fabric links drawn between them. Each
node is one actor whose body is a simple mesh; its color encodes role, and for
compute blades it additionally encodes health, so that degraded or offline nodes
are visible at a glance. Small device ``badges'' (LCD Panel) on each blade summarize its
processors, accelerators, memory, and storage. The visualizer also records a
lookup from node identifier to actor, which both the simulator and the
interactive layer reuse to resolve a name to a specific object in the world.

\subsection{Job and Hardware-Activity Simulation}
A simulation director (SimDirector) owns a virtual clock and an event schedule loaded from a
file. As the clock advances it starts and ends events, dispatching activity to
the affected blades through the identifier-to-actor registry. Blades animate
themselves: while at least one job is running on a blade it ``breathes'' and its
compute badges glow in the job's color (Fig.~\ref{fig:lcd}), returning to their idle appearance when
the last job ends. The director exposes transport controls (ready, play, pause,
stop, and finished states), adjustable time scaling, and replay, and it can
capture frames automatically to assemble time-lapse videos. Crucially for this
work, the director publishes the current virtual time and per-node activity,
which the interactive display samples.

\subsection{Navigation and the Command Shell}
The user explores the scene with a free-flying camera. Since the mouse-look
captures the pointer, \sysname also provides an in-application command shell (see red-outlined element in Fig.~\ref{fig:lcd}), a
small line-oriented interpreter overlaid on the view, for symbolic interaction.
Shell commands list and inspect nodes, search by role or status, drive the
simulation, frame the camera on a named scope, inject ad-hoc jobs, and
highlight or select nodes. This shell is the symbolic counterpart to direct
pointing, and, as described next, both feed the same selection state.

\subsection{Interactive Node Selection and the In-World LCD Panel}
The central new capability is the ability to \emph{choose} a node and see its
details in place. Selection is available through two paths that converge on a
single method so that there is exactly one ``selected” node and one reused
display. In the direct path, a left click casts a ray from the camera along its
forward direction. If the ray strikes a node's body, it selects
it; clicking empty space dismisses the display. A small crosshair is drawn at
the screen center so the pick is aimable. In the symbolic path, the shell's
select command resolves an identifier to a node or a rack, focuses the camera on
it, and triggers the same display. Only nodes and racks participate in this pick.

The display is an in-world panel, an ``LCD screen", rather than a
two-dimensional overlay, so that the information stays spatially associated with
its node. The panel is a single reused actor consisting of a translucent backing
quad and a text readout. When a node is selected, the panel is positioned beside
that node and made visible; while visible, it turns each frame to face the active
camera, so the readout remains legible from any viewing angle, and it rebuilds
its contents on a fixed sub-second cadence so that time-varying values update
smoothly. The backend uses a canvas or D3~\cite{bostock2011d3} update enabling reuse of visual analytics tools from a 2D layout as seen in Figure~\ref{fig:lcd}, which illustrates the panel attached to a selected
node. A video capture feature is built in to capture any trigger or error events as the time progresses,  Fig.~\ref{fig:sysoverview}. 

The text readout combines static and dynamic fields. From the data model, it shows the
node's identifier, role, and status, and its device specification. 

\subsection{Live Telemetry}
Production per-node telemetry is not yet connected, so the utilization and
temperature shown are \emph{synthesized} as a faithful placeholder, while the
running-job count is real state taken directly from the simulator. The synthesis
is deliberately tied to genuine activity so the numbers move believably: each
node's utilization is derived from its actual job load plus a smooth,
per-node-phased oscillation over virtual time, reading low when the node is idle
and high when it is busy, and pinned to zero when the node is offline.
Temperature tracks utilization with a small, stable per-node offset, so it rises
under load and settles when idle. Replacing the synthesized values
with recorded telemetry is a local change behind the same interface.
\section{Evaluation}
Our current evaluation establishes feasibility and characterizes
behavior rather than claiming a completed user study or performance campaign.

\subsection{Implementation Status and Methodology}
The platform and the new interactive display are implemented and built cleanly
for both the editor and standalone-game configurations of the engine on a
commodity Apple-silicon laptop. We verified end-to-end startup by launching the
application: the game mode loads, the scene is constructed from the
specification, the simulator advances and begins scheduled jobs, and the
frame-capture pipeline runs, all without errors on exit. Since the in-world
panel is instantiated only upon selection (an inherently interactive event),we
exercised it manually in a rendered session using both the click and the shell
paths.

\subsection{Qualitative Observations}
In interactive use, the two selection paths behave as intended and share a single
selected-node state, so pointing at a node and naming it in the shell are
interchangeable and never disagree. The panel appears beside the chosen node,
remains legible as the camera orbits because it billboards, and updates its live
fields continuously as the virtual clock advances; utilization and temperature
visibly rise when the simulator schedules jobs onto the node and relax when the
jobs end, and the running-job count matches the animated activity on the node.
The translucent backing preserves context: the selected node and its neighbors
remain visible through the panel, which is important when the point of a spatial
twin is to keep the surroundings in view.

\subsection{Runtime Behavior and Overhead}
The interactive features are designed to be inexpensive. Selection performs a
single ray query per click, and the reused panel is one lightweight actor with a
text component that rebuilds its string a few times per second rather than every
frame. The metric synthesis is closed-form per node or rack (CLI only) and evaluated only for the
one selected component on each refresh, so it contributes negligibly to frame time.
Rendering cost is dominated by the modeled subset of the cluster, which is
configurable. We defer a
rigorous frame-time and scalability study, including behavior when many nodes or
many simultaneous panels are shown, to future work.
\section{Discussion}
\label{sec:discussion}

\subsection{Limitations}
The system currently supports only a single selected node or rack and one
panel at a time, without aggregating status upward from blades to chassis,
and does not yet visualize inter-rack network topology. We have not
conducted a formal user study or a full-floor performance evaluation; these
bound our claims but not the architecture.

\subsection{Toward Real Telemetry and In-Situ Plots}
Our immediate roadmap addresses these limitations directly: visualizing
inter-rack network topology in the same object-model approach; introducing
per-node and per-rack telemetry read from recorded, timestamped data (power,
temperature, current) that the panel prefers over synthesized values when
available; and driving displayed values from the virtual clock, interpolating
each series so the panel replays recorded history and respects transport
controls while switching between inter-node and inter-rack readouts. Longer
term, we intend to aggregate health from blades up through chassis and racks,
support multiple simultaneous panels with level-of-detail management, and
conduct scalability and usability studies.

Taken together, this work closes the gap between the spatial and dynamic
views in an HPC digital twin: a user can point at, or name, any node and
read its state in place.

\section{Conclusion}

We presented \sysname, a parametric digital twin that renders a supercomputer as a navigable 3D scene, expanding a compact, declarative specification into typed components rather than an exact geometric replica. We used the Aurora supercomputer \cite{b3} as our first modeled machine. This abstraction-driven approach separates data, expansion, and presentation, so the same architecture can represent a different machine by editing a specification rather than rebuilding the visualization. Beyond static topology, \sysname layers in job-schedule simulation, an in-world telemetry overlay, and an interactive shell for querying, driving, and recording the cluster live, supporting both structural and operational understanding. Future work includes live telemetry, full-scale rendering via instanced meshes, and extending the event model to failures and thermal events. \textit{The code will be made available on publication}.

\section{Acknowledgments}
This research used resources of the Argonne Leadership Computing Facility, which is a U.S. Department of Energy Office of Science User Facility operated under contract DE-AC02-06CH11357. All authors were supported by the Office of Science, U.S. Department of Energy, under contract DE-AC02-06CH11357.

\bibliographystyle{IEEEtran}
\bibliography{digital-twin-references}

\end{document}